# Principle of Minimum Distance in Space of States as New Principle in Quantum Physics

D. B. Ion[1,2], M. L. D. Ion[3]

[1] IFIN-HH, Bucharest, P.O.Box MG-6, Magurele, Romania
[2] TH-Division, CERN, CH-1211 Geneva 23, Switzerland
[3] Faculty of Physics, Bucharest University, Bucharest, Romania

## Abstract

The mathematician Leonhard Euler (1707-1783) appears to have been a philosophical optimist having written: *"Since the fabric of universe is the most perfect and is the work of the most wise Creator, nothing whatsoever take place in this universe in which some relation of maximum or minimum does not appear. Wherefore, there is absolutely no doubt that every effect in universe can be explained as satisfactory from final causes themselves the aid of the method of Maxima and Minima, as can from the effective causes"*. Having in mind this kind of optimism in the papers [1-16] we introduced and investigated the possibility to construct a predictive analytic theory of the elementary particle interaction based on the principle of minimum distance in the space of quantum states (PMD-SQS). So, choosing the partial transition amplitudes as the system variational variables and the *distance in the space of the quantum states* as a measure of the system effectiveness, we obtained the results [1-16]. These results proved that the principle of minimum distance in space of quantum states (PMD-SQS) can be chosen as variational principle by which we can find the analytic expressions of the partial transition amplitudes. In this paper we present a description of hadron-hadron scattering via principle of minimum distance PMD-SQS when the distance in space of states is minimized with two directional constraints: $d\sigma/d\Omega(\pm 1) = fixed$. Then by using the available experimental (pion-nucleon and kaon-nucleon) phase shifts we obtained not only consistent experimental tests of the PMD-SQS optimality, but also strong experimental evidences for new principles in hadronic physics such as: *Principle of nonextensivity conjugation via the Riesz-Thorin relation (1/2p+1/2q=1)* and a new *Principle of limited uncertainty in nonextensive quantum physics*. The strong experimental evidence obtained here for the nonextensive statistical behavior of the $[J,\theta]$−quantum states in the pion-nucleon, kaon-nucleon and antikaon-nucleon scatterings can be interpreted as an indirect manifestation the presence of the *quarks and gluons as fundamental constituents* of the scattering system having the strong-coupling long-range regime required by the Quantum Chromodynamics.

## 1. Introduction

From historical point of view the earliest optimum principle was proposed by Heron of Alexandria (125 B.C.) in connection with the behaviour of light. Thus, Heron proved mathematically the following first genuine scientific minimum principle of physics: that *light travels between two points by shortest path*. In fact the Archimedean definition of a straight line as the shortest path between two points was an early expression of a variational principle, leading to the modern idea of a geodesic path. In fact in the same spirit, Hero of Alexandria explained the paths of reflected rays of light based on the principle of minimum distance (PMD), which Fermat (1657) reinterpreted as a principle of least time, Subsequently, Maupertuis and others developed this approach into a general principle of least action, applicable to mechanical as well as to optical phenomena. Of course, a more correct statement of these optimum principles is that systems evolve along stationary paths, which may be maximal, minimal, or neither (at an inflection point). Laws of mechanics were first formulated in terms of minimum principles. Optics and mechanics were brought together by a single minimum principle conceived by W. R. Hamilton. From Hamilton's single minimum principle could be obtained all the optical and mechanical laws then known. But the effort to find optimum principles has not been confined entirely to the exact sciences. In modern time the principles of optimum are extended to all sciences. So, there exists many minimum principle in action in all sciences, such as: *principle of minimum action, principle of minimum free energy, minimum charge, minimum entropy production, minimum Fischer information, minimum potential energy, minimum rate of energy dissipation, minimum dissipation, minimum of Chemical distance, minimum cross entropy, minimum complexity in evolution, minimum frustration, minimum sensitivity, etc.* So, a variety of generalizations of classical variational principles have appeared, and we shall not describe them here.

In the last decade, in the papers [1-16] we generalized these ideas to the space of transition amplitudes by introducing the *Principle of Minimum Distance in the Space of States* (PMD-SQS). These results can be grouped in four categories of results as follows: New optimum principles [1-5], New entropic uncertainty relations [6-10], Limited entropic uncertainty as a new principle of quantum physics [11,12,15], Theory of nonextensive quantum statistics and optimal entropic band [12-16]. To the above results we must add that, using the pion-nucleus (49 sets) and pion-nucleon (88 sets) experimental phase shifts we presented experimental evidences for *nonextensive quantum statistical behaviour* [12-16] in hadron-hadron and hadron-nucleus scattering.

In this paper, some new results on the optimal state analysis of hadron-hadron ($(\pi N, KN, \overline{K}N)$-scatterings, obtained by using the nonextensive quantum entropy [7-9] and principle of minimum distance in the space of quantum states (PMD-SQS) $\}$[1], are presented. Then, using $S_J(p), S_\theta(q), S_{J\theta}(p,q), \overline{S}_{J\theta}(p,q)$-Tsallis-like scattering entropies, the PMD-SQS-optimality as well as the nonextensive statistical behavior of the $[J], [\theta], [J,\theta]$ quantum systems of states produced in hadronic scatterings are investigated in an unified manner. A connection between optimal states obtained from the principle of minimum distance in the space of quantum states (PMD-SQS) [1] and the most stringent (MaxEnt) entropic bounds on Tsallis-like entropies for quantum scattering, is established. The nonextensivity indices p and q are determined from the experimental entropies by a fit with the optimal Tsallis-like entropies. In this way strong experimental evidences for the p-nonextensivities index in the range p=0.6 with q=p/(2p-1)=3, are evidentiated with high accuracy from the experimental data of the principal hadron-hadron scatterings.

## 2. Description of quantum scattering via principle of minimum distance in space of states

It is well known that any optimizing study [19] ideally involves three steps:

(I) The _description of the system,_ by which one should know, accurately and quantitatively, the essential variable of the system as well as how these system variables interact; (II) _Finding a unique measure of the system effectiveness_ expressible in terms of the system variables; (III) The _optimization_ by which one should chose those values of the system variables yielding optimum effectiveness.

In attempting to use the optimization theory for analyzing the interaction of elementary particles one can reverses the order of these three steps. Then, knowledge about the interacting system can be deduced by assuming that it behaves as to optimize some given measure of its effectiveness, and thus the behaviour of the system is completely specified by identifying the criterion of effectiveness and applying optimization to it. This approach is in fact known as describing the system in terms of an optimum principle.

Therefore, having in mind all successful results, obtained by using optimum principles in the entire history of physics from last two centuries, the principle of minimum distance was extended to the Hilbert space of the quantum states by choosing the partial transition amplitudes as fundamental physical quantities since they are labelled with good quantum numbers (such as charge, angular momentum, isospin, etc.). These fundamental physical quantities are chosen as the system variational variables while the distance in the Hilbert space of the quantum states is taken as measure of the system effectiveness expressed in terms of the system variables. Thus, the principle of minimum distance in the space of states (PMD-SQS) is chosen as variational principle by which one should obtain those values of the 'partial amplitudes yielding optimum effectiveness. Of course, the PMD-SQS-optimum principle can be formulated in a more general mathematical form by using the S-matrix theory of the strong interacting systems and reproducing kernel Hilbert spaces methods [18]. Then, a new analytic quantum physics can be developed in terms of the reproducing functions from the reproducing kernel Hilbert spaces (RKHS) of the transition amplitudes. In this new kind of analytic quantum physics the system variational variables will be the partial transition amplitudes introduced by the development of S-matrix elements in terms Fourier component implied by the fundamental symmetry of the quantum interacting system. Therefore, all the results will be expressed in terms of reproducing Kernel functions RKHS [18]. More concretely, applications to the PMD-SQS shown that this optimum principle can give a satisfactory account for diffraction scattering of any material particle (such as: electrons, neutrons, nuclei, atoms, molecules, etc.) as a universal optimal phenomenon, free of any (dualistic, wave or particle) interpretations.

In the concrete mathematical form, for the two body elastic scattering, the PMD-SQS can be reduced to the determination of the partial scattering amplitudes by solving giving minimization problems subject to some constraints imposed by interaction.

As example we chose here the case of the meson-nucleon scattering when the constraints are given just by fixing the differential cross sections in forward (x=+1) and backward (x=-1) directions.

So, first we present the basic definitions on the spin $(0^-1/2^+ \rightarrow 0^-1/2^+)$ scatterings

$$M(0^-) + N(1/2^+) \rightarrow M(0^-) + N(1/2^+) \qquad (1)$$

Therefore, let $f^{++}(x)$ and $f^{+-}(x)$, $x \in [-1,1]$ be the scattering helicity amplitudes of the meson-nucleon scattering process (see ref.[13]) written in terms of the partial helicities $f_{J-}$ sand $f_{J+}$ as follows

$$f_{++}(x) = \sum_{J=\frac{1}{2}}^{J_{\max}}\left(J + \frac{1}{2}\right)(f_{J-} + f_{J+})d^{J}_{\frac{1}{2}\frac{1}{2}}(x)$$

$$f_{+-}(x) = \sum_{J=\frac{1}{2}}^{J_{\max}}\left(J + \frac{1}{2}\right)(f_{J-} - f_{J+})d^{J}_{-\frac{1}{2}\frac{1}{2}}(x)$$

(2)

where the rotation functions are defined as

$$d^{J}_{\frac{1}{2}\frac{1}{2}}(x) = \frac{1}{l+1}\cdot\left[\frac{1+x}{2}\right]^{\frac{1}{2}}\left[\overset{\bullet}{P}_{l+1}(x) - \overset{\bullet}{P}_{l}(x)\right]$$

$$d^{J}_{-\frac{1}{2}\frac{1}{2}}(x) = \frac{1}{l+1}\cdot\left[\frac{1-x}{2}\right]^{\frac{1}{2}}\left[\overset{\bullet}{P}_{l+1}(x) + \overset{\bullet}{P}_{l}(x)\right]$$

(3)

where $P_l(x)$ are Legendre polinomials, $\overset{o}{P}_l(x) = \dfrac{d}{dx}P_l(x)$, x being the c.m. scattering angle. The normalisation of the helicity amplitudes $f^{++}(x)$ and $f^{+-}(x)$, is chosen such that the c.m. differential cross section is given by

$$\frac{d\sigma}{d\Omega}(x) = |f_{++}(x)|^2 + |f_{+-}(x)|^2$$

(4)

Then, the elastic integrated cross section is given by

$$\sigma_{el}/2\pi = \sum_{J=\frac{1}{2}}^{J_{\max}}(2J+1)\left(|f_{J+}|^2 + |f_{J-}|^2\right)$$

(5)

Now, let us consider the following two directional optimization problem:

$$\min\sum\left(j + \frac{1}{2}\right)\left[|f_{j+}|^2 + |f_{j-}|^2\right] \quad \text{when } \frac{d\sigma}{d\Omega}(+1) \text{ and } \frac{d\sigma}{d\Omega}(-1) \text{ are fixed}$$

(6)

where $\dfrac{d\sigma}{d\Omega}(\pm 1)$ are the forward and backward differential cross sections written in terms of partial amplitudes as follows

$$\frac{d\sigma}{d\Omega}(+1) = |f_{++}(+1)|^2 = |\sum_{J=\frac{1}{2}}^{J_{\max}}\left(J + \frac{1}{2}\right)(f_{J-} + f_{J+})|^2$$

(7)

$$\frac{d\sigma}{d\Omega}(-1) = |f_{+-}(-1)|^2 = |\sum_{J=\frac{1}{2}}^{J_{\max}}\left(J + \frac{1}{2}\right)(f_{J-} - f_{J+})|^2$$

(8)

Therefore, in this case the variational variables are partial helicity amplitudes $f_{J+}$ and $f_{J-}$ while the measure of the system effectiveness and the constraints are completely expressed in terms of these variational variables by eqs. (5)-(8).

We proved that the solution of this optimization problem is given by the following results :

$$f_{++}(x) = f_{++}(+1)\frac{K_{\frac{1}{2}\frac{1}{2}}(x,+1)}{K_{\frac{1}{2}\frac{1}{2}}(+1,+1)}, \quad f_{+-}^{o}(x) = f_{+-}(-1)\frac{K_{\frac{1}{2}-\frac{1}{2}}(x,-1)}{K_{\frac{1}{2}-\frac{1}{2}}(+1,-1)} \tag{7}$$

where the functions K(x,y) are the reproducing kernels [1-5] expressed in terms of the rotation function by

$$K_{\frac{1}{2}\frac{1}{2}}(\boldsymbol{x},\boldsymbol{y}) = \sum_{1/2}^{J_o}(j+\frac{1}{2})d_{\frac{1}{2}\frac{1}{2}}^{j}(\boldsymbol{x})d_{\frac{1}{2}\frac{1}{2}}^{j}(\boldsymbol{y}), \tag{8}$$

$$K_{\frac{1}{2}-\frac{1}{2}}(\boldsymbol{x},\boldsymbol{y}) = \sum_{1/2}^{J_o}(j+\frac{1}{2})d_{\frac{1}{2}-\frac{1}{2}}^{j}(\boldsymbol{x})d_{\frac{1}{2}-\frac{1}{2}}^{j}(\boldsymbol{y}), \tag{9}$$

while the optimal angular momentum is given by

$$\boldsymbol{J_0} = \sqrt{\frac{4\pi}{\sigma_{el}}\left[\frac{d\sigma}{d\Omega}(+1)+\frac{d\sigma}{d\Omega}(-1)\right]+\frac{1}{4}} - 1 \tag{10}$$

Now, let us consider the logarithmic slope b of the forward diffraction peak defined by

$$b = \frac{\boldsymbol{d}}{\boldsymbol{dt}}\left[\ln\frac{d\sigma}{dt}(\boldsymbol{s},\boldsymbol{t})\right]|_{t=0} \tag{11}$$

Then, using the definition of the rotation functions, from (7)-(11) we obtain the following optimal slope $b_o$

$$\boldsymbol{b_o} = \frac{\lambda^2}{4}\left[\frac{4\pi}{\sigma_{el}}\left(\frac{d\sigma}{d\Omega}(+1)+\frac{d\sigma}{d\Omega}(-1)\right)-1\right] \tag{12}$$

PMD-SQS-optimal predictions on the differential cross section $\frac{d\sigma_o}{d\Omega}(\boldsymbol{x})$ and also for the spin-polarization parameters $(P_o, R_o, A_o)$, are as follows.

$$\frac{d\sigma^{o\pm1}}{d\Omega}(x) = \frac{d\sigma}{d\Omega}(+1)\left[\frac{K_{\frac{1}{2}\frac{1}{2}}(x,+1)}{K_{\frac{1}{2}\frac{1}{2}}(+1,+1)}\right]^2 + \frac{d\sigma}{d\Omega}(-1)\left[\frac{K_{-\frac{1}{2}\frac{1}{2}}(x,-1)}{K_{-\frac{1}{2}\frac{1}{2}}(-1,-1)}\right]^2 \tag{13}$$

$$P_o\frac{d\sigma^{o\pm1}}{d\Omega}(x) = 2\sqrt{\frac{d\sigma}{d\Omega}(+1)}\sqrt{\frac{d\sigma}{d\Omega}(-1)}\left[\frac{K_{\frac{1}{2}\frac{1}{2}}(x,+1)}{K_{\frac{1}{2}\frac{1}{2}}(+1,+1)}\right]\left[\frac{K_{-\frac{1}{2}\frac{1}{2}}(x,-1)}{K_{-\frac{1}{2}\frac{1}{2}}(-1,-1)}\right]\sin\phi(x) \tag{14}$$

$$R_o\frac{d\sigma^{o\pm1}}{d\Omega}(x) = 2\sqrt{\frac{d\sigma}{d\Omega}(+1)}\sqrt{\frac{d\sigma}{d\Omega}(-1)}\left[\frac{K_{\frac{1}{2}\frac{1}{2}}(x,+1)}{K_{\frac{1}{2}\frac{1}{2}}(+1,+1)}\right]\left[\frac{K_{-\frac{1}{2}\frac{1}{2}}(x,-1)}{K_{-\frac{1}{2}\frac{1}{2}}(-1,-1)}\right]\cos\phi(x) \tag{15}$$

$$A_o \frac{d\sigma^{o\pm1}}{d\Omega}(x) = \frac{d\sigma}{d\Omega}(+1)\left[\frac{K_{\frac{1}{2}\frac{1}{2}}(x,+1)}{K_{\frac{1}{2}\frac{1}{2}}(+1,+1)}\right]^2 - \frac{d\sigma}{d\Omega}(-1)\left[\frac{K_{-\frac{1}{2}\frac{1}{2}}(x,-1)}{K_{-\frac{1}{2}\frac{1}{2}}(-1,-1)}\right]^2 \qquad (16)$$

$$\cos\phi(x) = \frac{\text{Re}\{[\boldsymbol{f}^{++}(+1)]^*\boldsymbol{f}^{+-}(-1)\}}{|\boldsymbol{f}^{++}(+1)||\boldsymbol{f}^{+-}(-1)|}, \quad \sin\phi(x) = \frac{\text{Im}\{[\boldsymbol{f}^{++}(+1)]^*\boldsymbol{f}^{+-}(-1)\}}{|\boldsymbol{f}^{++}(+1)||\boldsymbol{f}^{+-}(-1)|} \qquad (17)$$

Finally, it is important to note that all above results can be extended to the scattering of particles with arbitrary spins by using RKHS method presented in the papers [4,5], and also to particle production phenomena.

## 3. Quantum distances between two isospin channels in pion-nucleon scattering

The quantum distances between two isospin channels, in the Hilbert space of the hadron-hadron scattering amplitudes, are introduced in our paper [2]. Then, the isospin quantum distances, as well as their isospin bounds, are expressed in terms of the channel cross sections and spin polarisation parameters. The essential features as well as the energy behavior of the isospin quantum distances, for the pion-nucleon scattering, are presented for all (s,t,u)-channels. Therefore, let $[\vec{\Delta}_s(\Delta_s^{++},\Delta_s^{+-}),\vec{N}_s(N_s^{++},N_s^{+-})]$ and $[\vec{N}_u(N_u^{++},N_u^{+-}),\vec{\Delta}_u(\Delta_u^{++},\Delta_u^{+-})]$ be the (s,u) channel helicity $\pi N$-scattering amplitudes in the pure isospin (I=3/2,1/2)-scattering states . If the helicity amplitudes are normalized as in Eqs. (4)-(5). Then, the quantum distances are defined by the following relations:

$$D(\Delta_v,N_v) = \left[\left\|\vec{\Delta}_s\right\|^2 + \left\|\vec{N}_s\right\|^2 - 2|<\vec{N}_s,\vec{\Delta}_s>|\right]^{1/2} \text{ for v=s,u} \qquad (18)$$

$$\left\|\vec{\Delta}_v\right\|^2 = \int_{-1}^{+1}\frac{d\sigma_{3/2v}}{d\Omega}(x)dx = \sigma_{3/2v}/2\pi = \text{ and } \left\|\vec{N}_v\right\|^2 = \int_{-1}^{+1}\frac{d\sigma_{1/2v}}{d\Omega}(x)dx = \sigma_{1/2v}/2\pi \text{ for v=s,u} \qquad (19)$$

$$<\vec{N}_v,\vec{\Delta}_v> = \int_{-1}^{+1}\vec{N}_v.\vec{\Delta}_v dx = \int_{-1}^{+1}\left[N_{v++}^*\Delta_{v++} + N_{v+-}^*\Delta_{v+-}\right]dx \quad \text{for v=s,u} \qquad (20)$$

A similar definition is given for the isospin quantum distance $D(F_0,F_1)$ between the t-channel isospin states $I_t = 0,1$ described by the $\boldsymbol{F_0}$ and $\boldsymbol{F_1}$ vector amplitudes, respectively. Numerical results on the quantum distances $D(N_s,\Delta_s)$, $D(F_0,F_1)$ and $D(N_u,\Delta_u)$ as well as on their isospin bounds for the (s,t,u)-channels in the pion-nucleon scattering are presented in Fig. 1 by white small circles. All these results are obtained by using pion-nucleon phase shifs given in Ref. [20].

The estimation of the isospin quantum distances via their isospin bounds is more suitable since these bounds are expressed only in terms of the integrated cross sections [see Eqs. (41), (45) and (49) in Ref. [1]]. The quantum distances $D(N_s,\Delta_s)$ and $D(N_u,\Delta_u)$ attain their absolute maximum values of 1.86 fm at the position of $\Delta(1236)$-resonance. As is seen from Fig.1 the maximum value $D_{\max}(F_0,F_1) = 0.58\,fm$ of the isospin t-channels is also attained in the range of $\Delta(1236)$ resonance. In particular, at $P_{LAB}$=10 GeV/c, we have $0.01\,fm \le D(N,\Delta) \le 0.06\,fm$, for (s,u)-isospin channels, $0.247\,fm \leq D(F_0, F_1) \leq 0.265\,fm$ for t-channels. At high primary energies these distances for (s,u)-isospin channels are going to zero at least as $P_{LAB}^{-1/2}$. Moreover, the isospin conservation in the pion-nucleon scattering impose strong constraints on isospin quantum distances, not only in the $\Delta(1236)$ -resonance region but also at high energies where the isospin bounds on these distances are saturated. Consequently, a linear relation between average spin polarization parameters is expected to be observed (see Ref. in [2]), especially at energies higher than 2 GeV/c. The estimations of the isospin quantum distances via their isospin bounds are more suitable since these bounds are expressed only in terms of the integrated cross sections [see Eqs. (41), (45) and (49) in Ref. [1]].

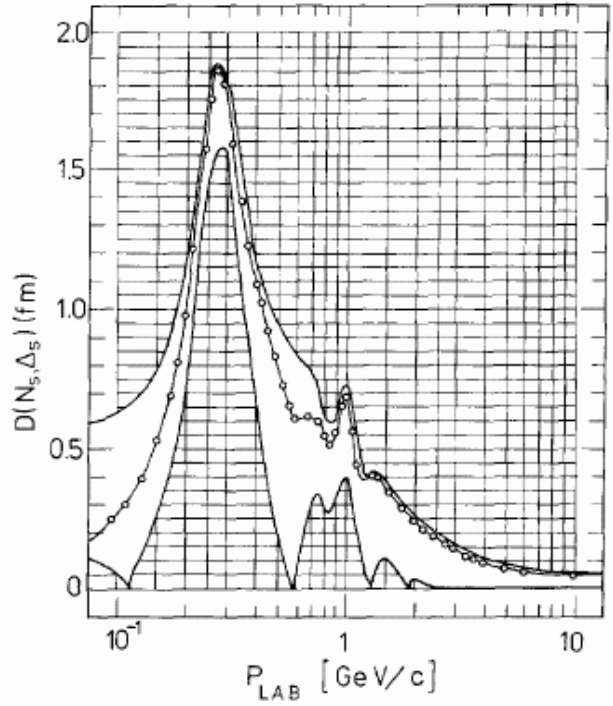

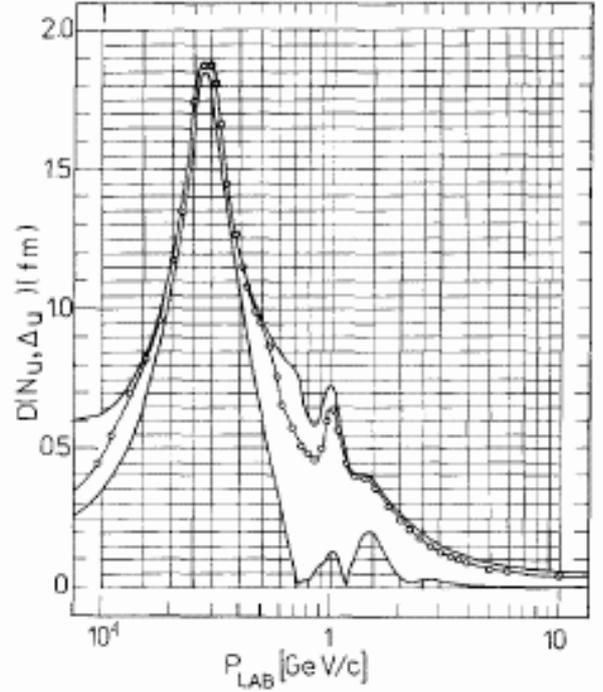

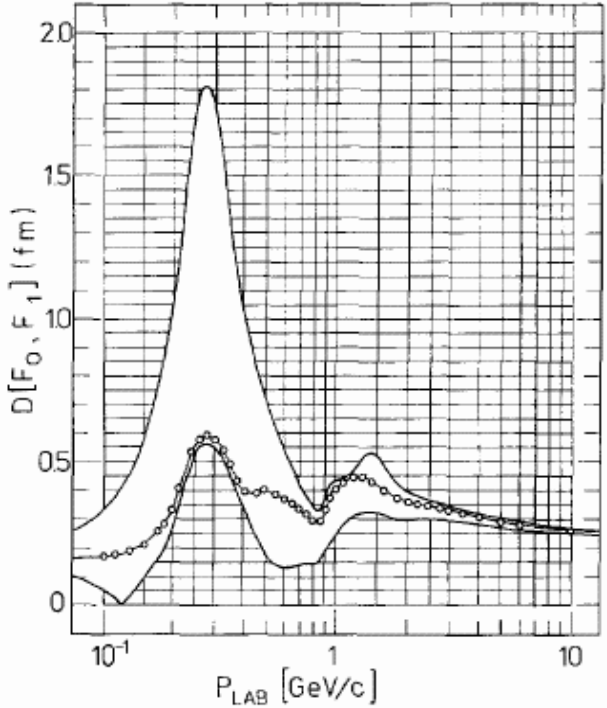

Fig.1 .The isospin quantum distance $D(\Delta_s, N_s)$ (o), $D(F_0, F_1)$ (o) and $D(\Delta_s, N_s)$ (o) and and their isospin bounds calculated by using Eqs. (38)-(41), (42)-(45) and (46)-(49) from Ref. [2] and pion-nucleon phase shifts from Ref. [20]. Due to the isospin invariance in the pion-nuclon scattering, the isospin quantum distances cannot take values outside of their isospin bounds in the hatched regions. (see the text).

## 4. Tests of the PMD-SQS optimality via logarithmic slope in the forward direction

Let **t** be usual transfer momentum transfer connected with the c.m scattering angle ( $x = \cos\theta$ ) by the usual relation: $t = -2q^2(1-x)$ where q is the c.m. momentum. Then the optimal logarithmic slope (12) is obtained in the following way

$$b_o = \frac{d}{dt} \ln\left[\frac{d\sigma^o}{dt}(s,t)\right]|_{t=0} = \frac{\hbar^2}{2}\frac{d}{dx}\left[\frac{d\sigma^o}{d\Omega}(x)\right]|_{x=0} =$$

$$= \frac{\hbar^2}{4}\left[(J_o+1)(J_o+2) - \frac{1}{4}\right] = \frac{\hbar^2}{4}\left[\frac{4\pi}{\sigma_{el}}\left(\frac{d\sigma}{d\Omega}(+1) + \frac{d\sigma}{d\Omega}(-1)\right) - 1\right]$$

Now, by the same procedure as in Ref. [1], we proved the following important optimal bound

$$b_o = \frac{\hbar^2}{4}\left[\frac{4\pi}{\sigma_{el}}\left(\frac{d\sigma}{d\Omega}(+1) + \frac{d\sigma}{d\Omega}(-1)\right) - 1\right] \le b \qquad (21)$$

The equality holds in (18) if and only if the helicity amplitudes are the optimal PMD-SQS solutions (7)-(10).

Therefore, for an experimental test of the PMD-SQS optimal result (7)-(10) the numerical valus of the experimental and optimal slopes are calculated from the experimental phase shifts (EPS) solutions [20], and [32] and also directly from the experimental data [25-35]. The results are displayed in Figs. 2-4. In the case of pion-nucleon scattering for $P_{LAB} \ge 3GeV/c$ the experimental data on $[b_{exp}, \frac{d\sigma}{d\Omega}, \sigma_{el}]$ are collected mainly from the original fits of Refs.[25]-[29]. To these data we added some values of b from Lasinski et al [33] and also calculate directly from the phase shifts (EPS) [20]. In the low laboratory momenta $P_{LAB} \le 2GeV/c$, the slope parameters $b_{exp}$ and $b_o$ are determined from the EPS-solutions [20]. The experimental data for $K^{\pm}P$ − scatterings are collected from the original fits [25],[27],[29],[31],[32] to which we added those pairs calculated from the experimental phase shifts (EPS) solutions of [32]. The experimental data for $[PP, \overline{P}P]$ − scatterings are obtained from Refs [33]-[35].

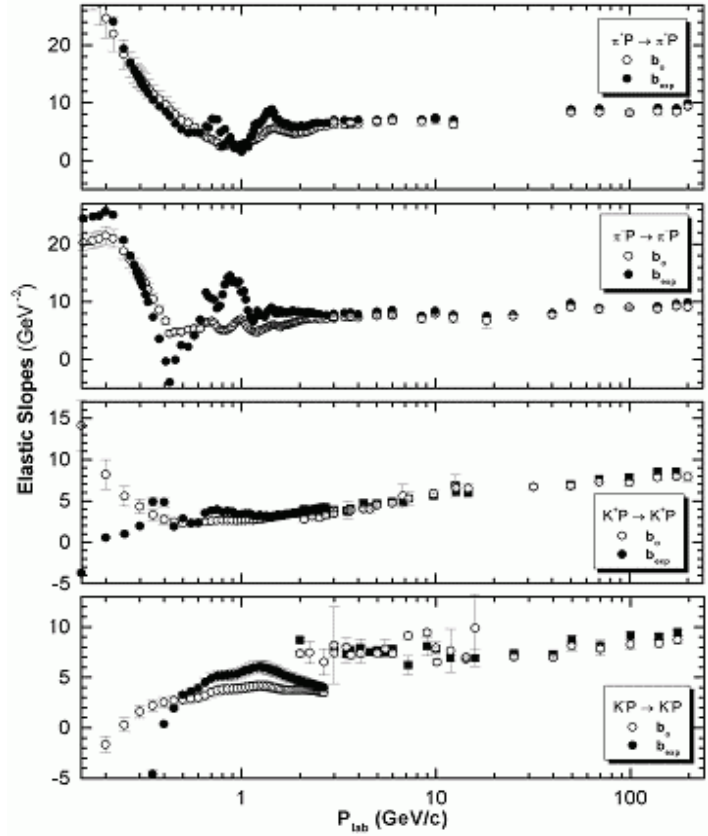

Fig. 2a: The experimental values of the logarithmic slopes $b_{\exp}$ (black circles) for meson-nucleon scatterings are compared with optimal state predictions $b_o$ (12) (white circles) (see the text)

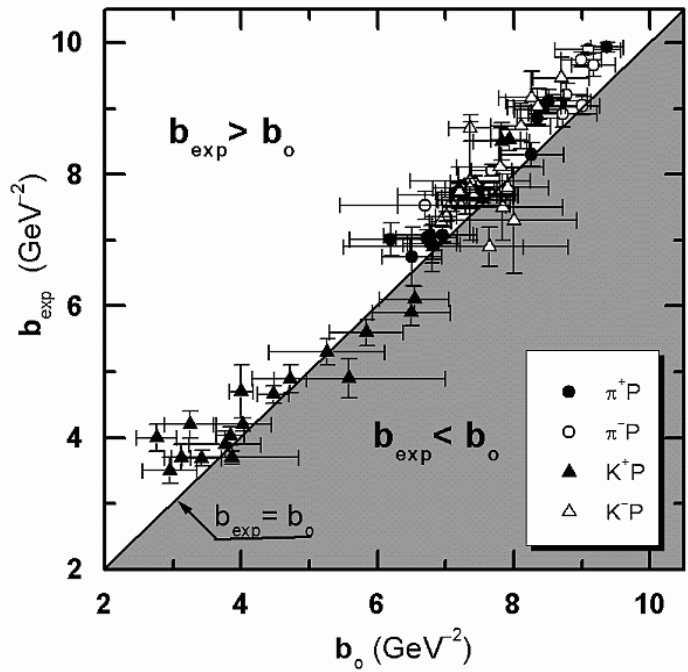

Fig. 2b: The experimental values of the logarithmic slopes $b_{\exp}$ for $[\pi^{\pm}P, K^{\pm}P]-$scatterings for $p_{LAB} \geq 2$ GeV/c are compared with the optimal state predictions $b_o$ (22) (see Figs. 2a,b).

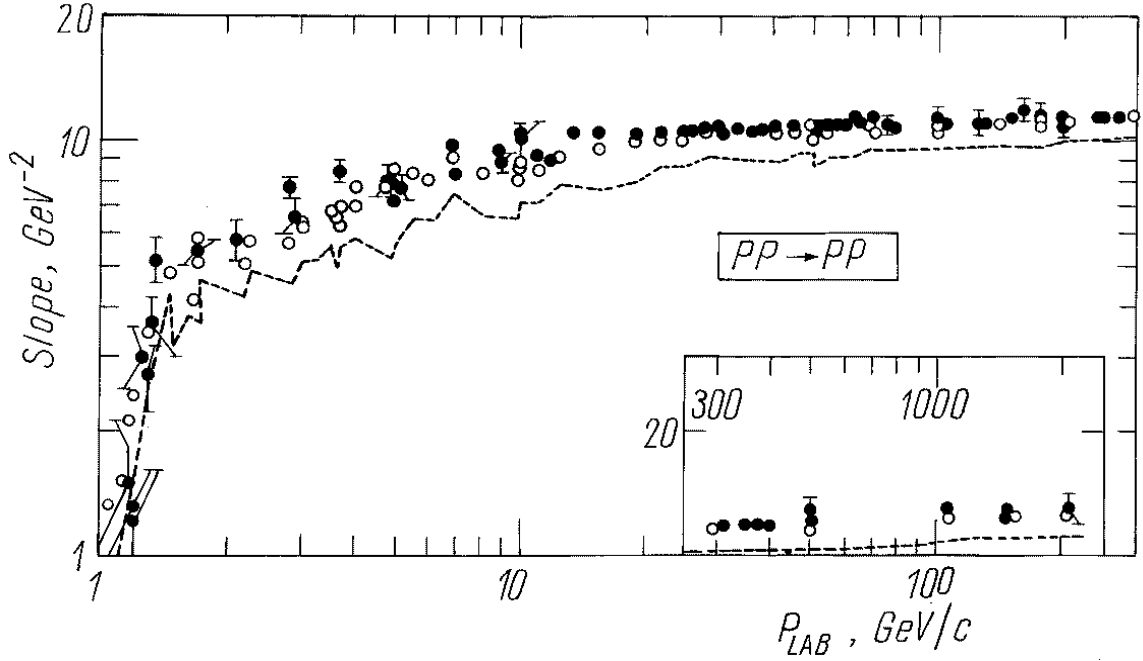

Fig. 3: The experimental values of the logarithmic slopes $b_{exp}$ for proton-proton scatterings (black circles)
are compared with the optimal state predictions $b_o$ (22) (white circles) (see the text). Dashed curve
correspond to an estimation of the Martin-MacDowell unitarity bound (25).

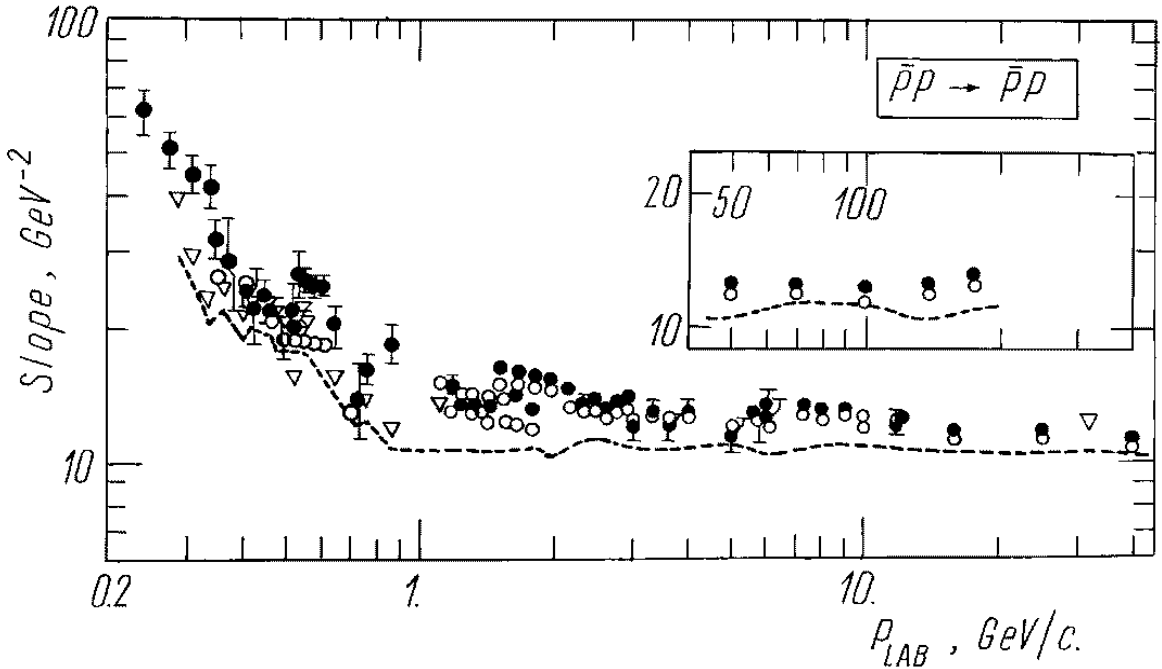

Fig. 4: The experimental values of the logarithmic slopes $b_{exp}$ for $\overline{P}P$ scatterings (blacke circles)
are compared with the optimal state predictions $b_o$.(22) (white circles and white triangles).
Dashed curve correspond to an estimation of the Martin-MacDowell unitarity bound (25).

Now, it is important to note that the optimal inequality (21) includes in a more general and exact
form the unitarity bounds derived by Martin [21] and Martin-Mac Dowell [22] and Ion [1,23,24].

Indeed, since $\dfrac{d\sigma}{d\Omega}(\pm 1) \geq 0$, from Eq. (21) it is easy to obtain

$$b_{o+1} = \frac{\lambda^2}{4}\left[\frac{4\pi}{\sigma_{el}}\frac{d\sigma}{d\Omega}(+1)-1\right] \leq b \quad \text{(proved by D.B.Ion [1])} \tag{22}$$

$$b_{o-1} = \frac{\lambda^2}{4}\left[\frac{4\pi}{\sigma_{el}}\frac{d\sigma}{d\Omega}(-1)-1\right] \leq b \quad \text{(proved by in Ref. [5,24])} \tag{23}$$

$$\frac{\lambda^2}{4}\left[\frac{\sigma_T^2}{4\pi\lambda^2\sigma_{el}}-1\right] \leq b, \text{ (D.B. Ion bound Refs. [5,23])} \tag{24}$$

$$\frac{2\lambda^2}{9}\left[\frac{\sigma_T^2}{4\pi\lambda^2\sigma_{el}}-1\right] \leq b_A, \text{ (Martin MacDowell bound [22])} \tag{25}$$

$$\frac{\lambda^2}{4}\left[\frac{\sigma_T}{4\pi\lambda^2}-1\right] \leq b_A \quad \text{(Martin-bound [21])} \tag{26}$$

For comparisons with optimal slopes in Figs. 3-4 the estimation of the Martin-MacDowell bound are also presented by dashed curves. Therefore, from Figs. 3-5, we conclude that the available experimental data on the differential cross sections for the forward peak of all [meson-nucleon, nucleon-nucleon and antinucleon-nucleon scatterings, at all energies are described in an unified way by optimal logarithmic slope $b_o$ (13).

## 5. Scaling of hadron-hadron scattering via PMD-SQS-optimality.

The theory of the optimal states from the reproducing kernel Hilbert space (RKHS) of the scattering amplitudes is developed in the papers [1,3-5]. Then, two important physical laws of hadron-hadron scattering the <u>scaling of angular distributions and S-channel helicity conservations</u> are derived via optimal state dominance in our papers [1,5] by using the PMD-SQS minimum principle. Now, it is important to note that all the available experimental data on the differential cross sections for the forward peak of all [meson-nucleon, nucleon-nucleon and antinucleon-nucleon scatterings, at all energies higher than 2 GeV, can be described in an unified way by the following optimal scaling functions

$$f(s,t)) = f(\tau) = \frac{d\sigma}{dt}(s,t)/\frac{d\sigma}{dt}(s,0), \quad \tau = 2\sqrt{|t|b_o(s)} \tag{27}$$

$$f^o(\tau) = \frac{d\sigma^o}{d\Omega}(x)/\frac{d\sigma}{d\Omega}(+1) = \left[\frac{2J(\tau)}{\tau}\right]^2 \quad \tau = 2[|t|b_o]^{1/2} \text{ for small } \tau \tag{28}$$

The experimental values of the scaling function, obtained by using Eqs. (27) and the experimental data [35,47-49], are compared with the theoretical predictions (28) for $[\pi^{\pm}P, K^{\pm}, PP, \overline{P}P]$-scatterings. These results allow us to conclude that a principle of minimum distance in the space of states can be a good candidate for an optimum principle in the theory of hadronic interactions.

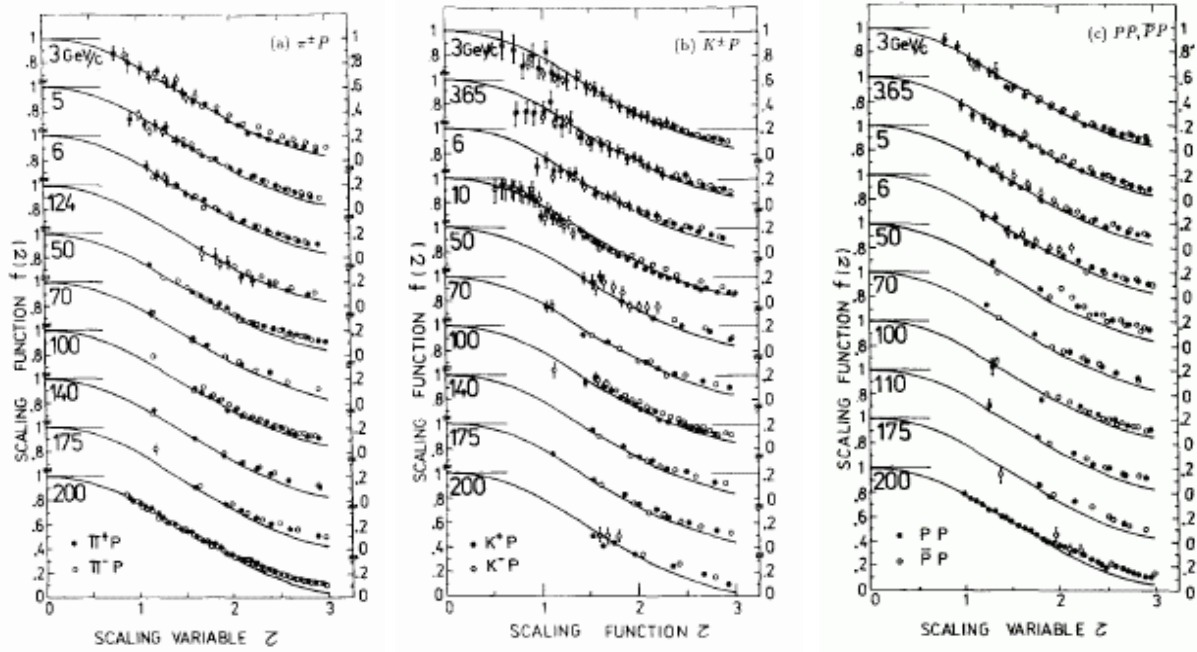

Fig 5. The experimental values of the scaling function , obtained by using Eqs. (27) and the experimental data [35,47-49],are compared with the theoretical predictions (28) for $[\pi^{\pm}P, K^{\pm}, PP, \overline{P}P]$-scatterings.

Next, it is easy to prove that the results (27) include in a more general and exact form the scaling variables $|t|\sigma_T^2/\sigma_{el}$ and $|t|\sigma_T$ introduced in Refs.[36]-[39] (see also [40]).

Indeed, as early as 1970 Singh and Roy [36] established that an upper bound for the absorptive ratio scale in the variable $t\sigma_T^2/\sigma_{el}$ for small |t|. Cornille and Marin [37] have examined the rigorous foundation of scaling properties of the differential cross section-ratio (see Eq. (27)) in the context of diffraction scattering. They have proved scaling in the variable $t\sigma_T^2/\sigma_{el}$ for the case when $\sigma_T/(\ln s)^2$ in the asymptotic. Moreover, Cornille [38] has examined different conditions of scaling and shown that one of of the scaling variable at asymptotic energies can be tb(s), (s,t,u) being the usual Mandelstam variables.

## 6. The PMD-SQS optimality and nonextensive statistical behaviour in hadronic scattering

In the last time there is an increasing interest in the foundation of a new statistical theory [52,53] valid for the nonextensive statistical systems which exhibit some relevant long range interactions, the memory effects or multifractal structures. It is important to mention here that the Tsallis nonextensive statistical formalism [53] already has been successfully applied to a large variety of phenomena such as (see Ref. [54]): Levy-like and correlated anomalous diffusions, turbulence in electron plasma, self-graviting systems, cosmology, galaxy clusters, motion of Hydra viridissima, classical and quantum chaos, quantum entanglement, reassociation in folded proteins, superstatistics, economics, linguistic, etc. Here, is worth to mention the recent applications of nonextensive statistics to nuclear and high-energy particle physics, namely: electron-positron annihilations [55,56], quark-qluon plasma [57], hadronic collisions [6-16,58], nuclear collisions [59], and solar neutrinos [60,61]. In general, in hadronic quantum systems the nonextensivity is expected to emerge due to the presence of the quarks and gluons having a strong cupling long-range regime of quantum chromodynamics. Hence, we think that the pion-nucleon, kaon-nucleon, antikaon-nucleon, as well as the pion-nucleus quantum scatterind systems are the most suitable for such investigations since in these cases the experimental nonextensive entropies can be obtained with high accuracy from the available phase shifts analyses [20]-[32].

## 6.1. *J-nonextensive statistics for the quantum scattering states*

We define two kind of Tsallis-like scattering entropies. One of them, namely $S_J(p)$, $p \in R$ is special dedicated to the investigation of the nonextensive statistical behavior of the angular momentum quantum states can be defined by

$$S_J(p) = \frac{1}{p-1}\left[1 - \sum_{J=1/2}^{J_{max}}(2j+1)p_j^p\right] \leq S_J^{o1}(p) \ , \quad p \in R \ ,$$

(29)

where the probability distribution are given by

$$p_j = \frac{|f_{j+}|^2 + |f_{j-}|^2}{\sum_{1/2}^{J_{max}}(2j+1)(|f_{j+}|^2 + |f_{j-}|^2)} \ , \quad \sum_{1/2}^{J_{max}}(2j+1)p_j = 1$$

(30)

## 6.2. *θ-nonextensive statistics for the quantum scattering states*

In similar way, for the $\theta$-scattering states considered as statistical canonical ensemble, we can investigate their nonextensive statistical behavior by using an angular Tsallis-like scattering entropy $S_\theta(q)$ defined as

$$S_\theta(q) = \frac{1}{q-1}\left[1 - \int_{-1}^{+1}dxP^q(x)dx\right] \ , \ q \in R$$

(31)

where the density of probality $P(x)$ is defined as follows

$$P(x) = \frac{2\pi}{\sigma_{el}}\frac{d\sigma}{d\Omega}(x) \quad , \quad \int_{-1}^{+1}dxP(x)dx = 1$$

(32)

with $\frac{d\sigma}{d\Omega}(x)$ and $\sigma_{el}$ are defined by Eqs.(4)-(5) and (6)-(7).

The above Tsallis-like entropies posses two important properties. First, in the limit $p \to 1$ and $q \to 1$, the Boltzman-Gibbs kind of entropies are recovered

$$\lim_{q \to 1} S_\theta(q) = -\int dxP(x)\ln P(x) \quad \text{and} \quad \lim_{p \to 1} S_J(p) = -\sum_{1/2}^{J_{max}}(2j+1)p_j \ln p_j$$

(33)

and

$$S_{A+B}(k) = S_A + S_B + (1-k)S_A(k)S_B(k) \quad \text{for } k = p, q \in R$$

(34)

for any independent subsystems ($p_{A+B} = p_A.p_B$). Hence, each of the indices $p \neq 1, q \neq 1$ from the definitions (29) and (31) can be interpreted as measuring the degree of nonextensivity.

### 6.3. $[J\theta]$-nonextensive statistics for the quantum scattering states

Also, using the same distribution (30) and (32), we define the combined $S_{J\theta}(p,q)$ entropies

$$\overline{S}_{J\theta}(p,q) = \frac{1}{p-1}\left[1 - \sum_{J=1/2}^{J_{max}}(2j+1)p_j^p\int_{-1}^{+1}dxP^q(x)dx\right], \ p \neq q \in R \tag{35}$$

$$\overline{S}_{J\theta}(q,q) = S_{J\theta}(q) \tag{36}$$

with the properties

$$S_{J\theta}(q) = S_J(q) + S_\theta(q) + (1-q)S_J(q)S_\theta(q) \tag{37}$$

$$\overline{S}_{J\theta}(p,q) = S_J(p) + \overline{S}_\theta(p) + (1-p)S_J(p)\overline{S}_\theta(p), \text{ where } \overline{S}_\theta(p) \equiv \frac{q-1}{p-1}S_\theta(q), \ p \neq q \in R \tag{38}$$

### 6.3. *The equilibrium distributions for the [J], [$\theta$] and [$J\theta$] quantum systems of states*

We next consider the maximum-entropy (MaxEnt) problem

$$\max\left[S_J(p), S_\theta(q), \overline{S}_{J\theta}(p,q)\right] \text{ when } \sigma_{el}, \frac{d\sigma}{d\Omega}(+1), \text{and } \frac{d\sigma}{d\Omega}(-1), \text{ are fixed} \tag{39}$$

as criterion for the determination of the *equilibrium distributions* for the quantum states produced from the meson-nucleon scattering. The *equilibrium distributions* $\{p_j^{me}\}$, $P^{me}(x)$, as well as the optimal scattering entropies for the quantum scattering of the spineless particles were obtained in Ref. [8-9]. For the *J-quantum states*, and *$\theta$ - quantum states* in the meson-nucleon scattering case these distributions are given by:

$$p_j^{me} = p_j^o = \left[(J_o+1)^2 - 1/4\right]^{-1} \text{ for all } 1/2 \leq j \leq J_o \text{ and } p_j = 0 \text{ for all } j \geq J_o + 1 \tag{40}$$

$$P^{me}(x) = P^o(x) = \frac{2\pi}{\sigma_{el}}\frac{d\sigma^o}{d\Omega}(x) \tag{41}$$

with $J_o$ and $\frac{d\sigma^o}{d\Omega}(x)$ given by Eqs. (10) and (13), respectively.

Indeed, solving problem (33) via Lagrange multipliers we get that the singular solution $\lambda_0 = 0$ exists and is just given by the $\left[S_J^o(p), S_\theta^o(q), S_{J\theta}^o(p), \overline{S}_{J\theta}^o(p,q)\right]$-optimal entropies corresponding to the PMD-SQS-optimal state (7)-(10). Therefore, the solution of problem (39) is given by

$$S_\theta(q) \leq S_\theta^o(q) = \frac{1}{q-1}\left[1 - \int_{-1}^{+1}\left[P^o(x)\right]^q dx\right], \ P^o(x) = \frac{2\pi}{\sigma_{el}}\frac{d\sigma^o}{d\Omega}(x) \text{ [see Eq. (13)]} \tag{42}$$

$$S_J(p) \leq S_J^o(q) = \frac{1}{p-1}\left[1 - \sum_{1/2}^{J_o}(2j+1)\left[p_j^o\right]^p\right], \ p_j^o = \frac{1}{(J_o+1)^2 - 1/4} \text{ for } 1/2 \leq j \leq J_o \tag{43}$$

$$\overline{S}_{J\theta}(p,q) \le \overline{S}^{\,o}_{J\theta}(p,q) = \frac{1}{p-1}\left[1 - \sum_{1/2}^{J_o}(2j+1)\left[p_j^{\,o}\right]^p \int_{-1}^{+1}\left[P^o(x)\right]^q dx\right]$$

<div align="right">(44)</div>

The equality holds in (42)-(44) if and only if the helicity amplitudes are the optimal PMD-SQS solutions (7)-(10).

Now, for a systematic experimental investigation of the saturation of the optimality limits in hadron-hadron scattering we used the available experimental phase-shifts [20,32] to solve the following important problems: (i) To reconstruct the experimental pion-nucleon, kaon-nucleon and antikaon-nucleon scattering amplitudes; (ii) To obtain numerical values of the experimental scattering entropies $S_J(p)$, $S_\theta(q)$ from the reconstructed amplitudes; (iii) To obtain the numerical values of the optimal $J_o$ from experimental scattering amplitudes, (iv) To calculate the numerical values for the PMD-SQS-optimal entropies $S_J^o(p)$, $S_\theta^o(q)$. So, these results presented in Fig. 6 are compared with the results of the PMD-SQS -optimal model from section 2. The grey regions from Figs. 6-7 are obtained by assuming a minimum error of ($\Delta J_o = \pm 1, \Delta L_o = \pm 1,$) in estimation of the optimal angular momenta ($J_o, L_o$), respectively.

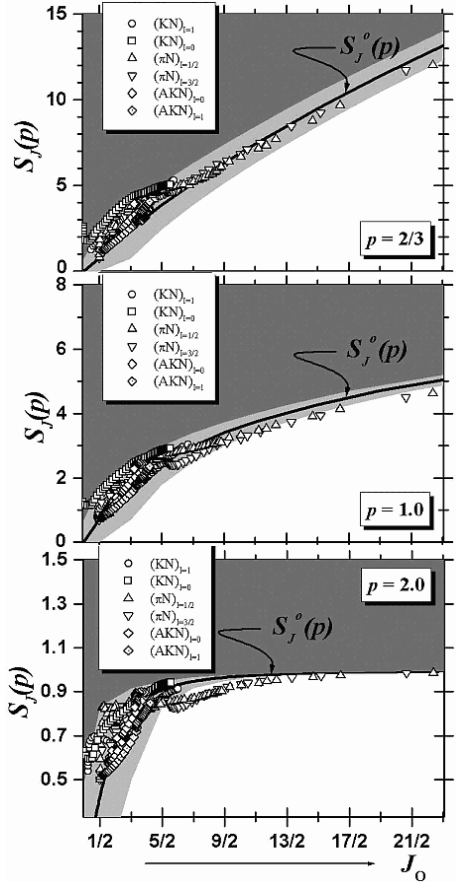

Fig. 6a: The experimental values of the meson-nucleon scattering entropies $S_\theta(p)$ (29), are compared with their optimal PMD-SQS upper bounds $S_J^0(p)$ (43) (solid curves) for the entropic nonextensivities p=2/3,1 and 2, respectively. The numerical values for $S_J(p)$ and $S_J^0(p)$ are obtained by using the available phase shifts analyses [20] and [32]. The grey region around the optimal entropies are obtained by assuming an error of $\Delta J_o = \pm 1$ in the estimation of optimal angular momentum $J_o$.

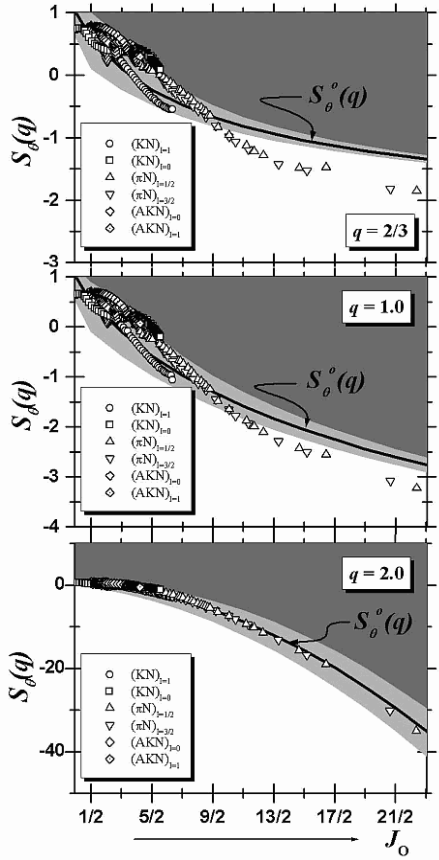

Fig. 6b: The experimental values of the meson-nucleon scattering entropies $S_\theta(q)$ (31), are compared with their optimal PMD-SQS upper bounds $S_\theta(q)$ (42) (solid curves) for the entropic nonextensivities p=2/3,1 and 2, respectively. The numerical values for $S_\theta(p)$ and $S_\theta^0(p)$ are obtained by using the available phase shifts analyses [20] and [32]. The grey region around the optimal entropies are obtained by assuming an error of $\Delta J_o = \pm 1$ in the estimation of optimal angular momentum $J_o$.

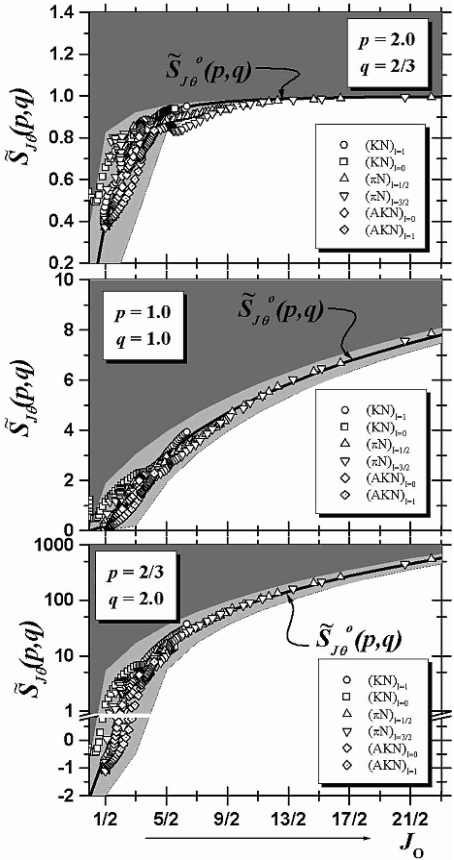

Fig. 6c: The experimental values of the meson-nucleon scattering entropies $\overline{S}_{J\theta}(p,q)$ (35), are compared with their optimal PMD-SQS upper bounds $S_\theta(q)$ (44) (solid curves) for the entropic nonextensivities p=2/3,1 and 2, respectively. The numerical values for $\overline{S}_{J\theta}(p,q)$ and $S_\theta^0(p)$ are obtained by using the available phase shifts analyses [20] and [32]. The grey region around the optimal entropies are obtained by aassuming an error of $\Delta J_o = \pm 1$ in the estimation of optimal angular momentum $J_o$.

From Fig. 6a we see that the experimental values of the scattering entropies $S_J(p)$ are well described with high accuracy by the optimal values $S_J^o(p)$ only for the statistical non-extensivitiy p=2/3 while a significant discrepancy is observed for p=1 and p=2. From Fig. 6b we see that the saturation of the optimality limit $S_\theta^o(q)$ is evidentiated for the values of $S_\theta(q)$ entropy corresponding to the conjugated nonextensivity q=2. Significant departures from the optimal entropy $S_\theta^o$ (solid line) is observed for nonextensivities q=2/3 and q=1, respectively.

Consequently, the entropies $\overline{S}_{J\theta}(p,q)$ are reproduced with high accuracy only for the pairs of nonextensivities $(p,q)=(2/3,2)$ and (1,1) (see Fig. 6c).

In particular, for the scattering of spinless particles we proved [9] the following entropic optimal bounds

$$S_\theta(q) \le S_\theta^{o1}(q) = \frac{1}{q-1}\left[1 - \int_{-1}^{+1}\left[\frac{[K(x,1)]^2}{K(1,1)}\right]^q dx\right] \quad \text{(see Ref. [9])} \tag{45}$$

$$S_L(q) \le S_L^{o1}(q) = \frac{1}{q-1}\left[1 - [L_o+1]^{2(1-q)}\right] \tag{46}$$

$$\overline{S}_{J\theta}(p,q) \le \overline{S}_{J\theta}^{o1}(p,q) = \frac{1}{q-1}\left[1 - [L_o+1]^{2(1-p)}\left[1 - \int_{-1}^{+1}\left[\frac{[K(x,1)]^2}{K(1,1)}\right]^q dx\right]\right] \tag{47}$$

$$K(x,1) = \frac{1}{2}\sum_0^{L_o}(2l+1)P_l(x) \quad \text{and} \quad 2K(1,1) = (L_o+1)^2 \tag{48}$$

So, the problem to find an upper bound for all Tsallis-like scattering entropies (45)-(48) when the elastic integrated cross section $\sigma_{el}$ and the forward differential cross section $d\sigma/d\Omega$ (1) are fixed is completely solved in terms of the optimal entropies obtained from PMD-SQS. The results on experimental tests of these optimal upper bounds are presented for both extensive and nonextensive statistics cases in Fig. 2 by using 49 sets of the available pion-nucleus (He, C, O, and Ca) phase shifts.

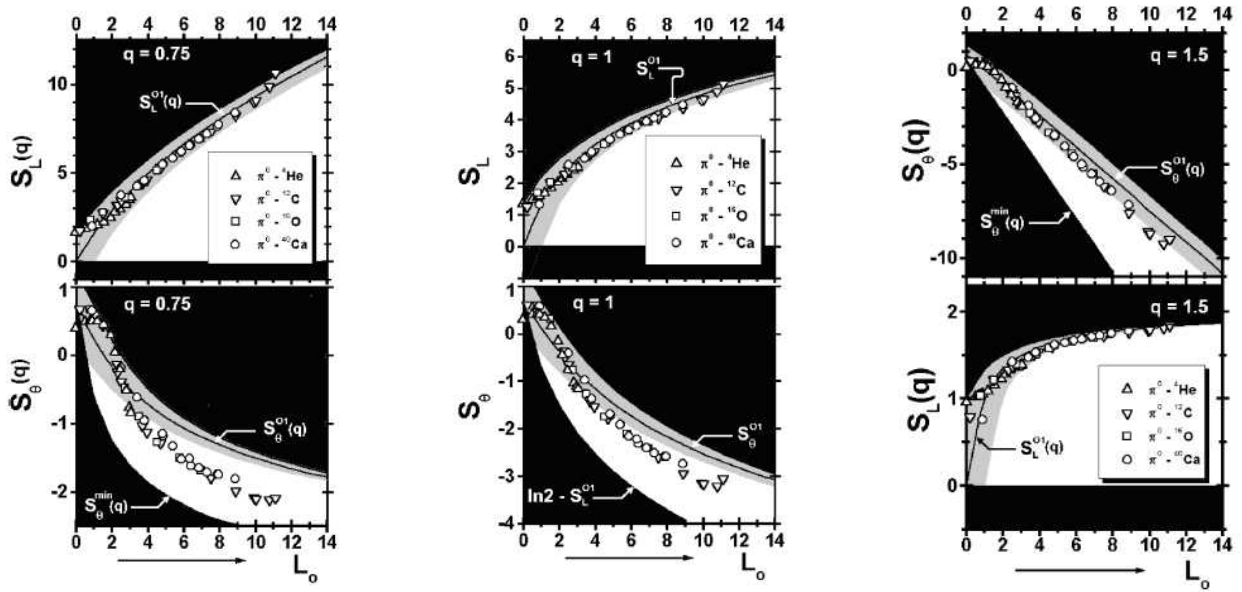

Fig 7. The experimental values of the scattering entropies $S_\theta(q)$ and $S_L(q)$, are compared with the theoretical optimal state predictions(full curve) for pion-nucleus (He, C, O, and Ca) scatterings (see the paper [9] for details).

## 7. Evidences for $(\frac{1}{2p} + \frac{1}{2q})$ – nonextensivity conjugation in quantum nonextensive scattering

A natural but fundamental question was addressed in Refs. [9-11], namely, what kind correlation (if it exists) is expected to be observed between the nonextensivity indices p and q corresponding to the (p,J)-nonextensive statistics described by $S_J(p)$-scattering entropy and (q, $\theta$)-nonextensive statistics described by $S_\theta(q)$-scattering entropy? In order to get an consistent mathematical answer to this question we first present the following important remarks.

<u>Remark 1:</u> Any Tsallis-like entropy $S_J(p)$ (29)-(30) can be written in the following equivalent form

$$S_J(p) = \frac{1}{p-1}\left[1 - \left(\left\|\varphi_j\right\|_{2p}\right)^{\frac{1}{2p}}\right], \qquad p \in R \tag{49}$$

$$\varphi_{j\pm} = \frac{f_{j\pm}}{\sqrt{\sum_{1/2}^{J_{max}}(2j+1)\left[\left|f_{j+}\right|^2 + \left|f_{j-}\right|^2\right]}}, \quad p_j = \left|\varphi_{j+}\right|^2 + \left|\varphi_{j-}\right|^2, \quad \{\varphi_j = [\varphi_{j+}, \varphi_{j-}]\} \in L_{2p} \tag{50}$$

$$\left\|\varphi_j\right\|_{2p} = \left[\sum_{1/2}^{J_{max}}(2j+1)p_j\right]^{\frac{1}{2p}} = \left[1 + (1-p)S_J(p)\right]^{\frac{1}{2p}} \tag{51}$$

and consequently the nonextensivity index p can be interprted as being directly connected with the dimension 2p of the Hilbert space $L_{2p}$ of the vector valued functions $\varphi_j = [\varphi_{j+}, \varphi_{j-}]$

<u>Remark 2</u>: Any scattering entropy $S_\theta(q)$ (31)-(32) can be written in the following equivalent form

$$S_\theta(q) = \frac{1}{q-1}\left[1 - \left(\left\|\phi\right\|_{2q}\right)^{2q}\right], \quad q \in R \tag{52}$$

$$\phi_{+\pm}(x) = \frac{f_{+\pm}(x)}{\sqrt{\int_{-1}^{+1}dx\left[\left|f_{++}(x)\right|^2 + \left|f_{+-}(x)\right|^2\right]}}, \quad P(x) = \left|\phi_{++}(x)\right|^2 + \left|\phi_{+-}(x)\right|^2, \quad \{\phi(x) = [\phi_{++}(x), \phi_{+-}(x)]\} \in L_{2q} \tag{53}$$

$$\left\|\phi(x)\right\|_{2p} = \left[\int_{-1}^{+1}dxP^q(x)\right]^{\frac{1}{2q}} = \left[1 + (1-q)S_\theta(q)\right]^{\frac{1}{2q}} \tag{54}$$

and consequently the nonextensivity index q can be interpreted as being directly connected with the dimension 2q of the Hilbert space $L_{2q}$ of the vector valued functions $\phi = [\phi_{++}(x), \phi_{+-}(x)]$.

**Riesz-Thorin (1/2p+1/2q)-correlation:** *If the Fourier transform defined by Eqs. (2)-(3) is a <u>bounded map</u> $T : L_{2p} \to L_{2q}$, then the nonextensivity indices (p,q) of the J -statistics described by the Tsallis-like entropy $S_J(p)$ and $\theta$ – statistics described by the scattering entropy $S_\theta(q)$, must be correlated via the Riesz-Thorin relation*

$$\frac{1}{2p} + \frac{1}{2q} = 1 \quad \text{or} \quad q = \frac{p}{2q-1} \tag{55}$$

*while the norm estimate of this map is given by*

$$M = \frac{\|Tf\|_{2q}}{\|f\|_{2p}} \leq 2^{\frac{p}{2q-1}}$$

(56)

Then, the results (55)-(56) can be obtained as a direct consequence of the Riesz-Thorin interpolation theorem extended to the vector-valued functions (see Ref. [15] for a detailed proof). Next, the nonextensivity indices $p$ and $q$ are determined [12] from the experimental entropies by a fit with the optimal entropies $[S_J^{o1}(p), S_0^{o1}(q)]$ obtained from the *principle of minimum distance in the space of states*. In this way strong experimental evidences for the $p$-nonextensivities in the range $0.5 \lesssim p \lesssim 0.6$ with $q=p/(2p-1)>3$ are obtained [with high accuracy (CL>99%)] from the experimental data of pion– nucleon and pion–nucleus scatterings

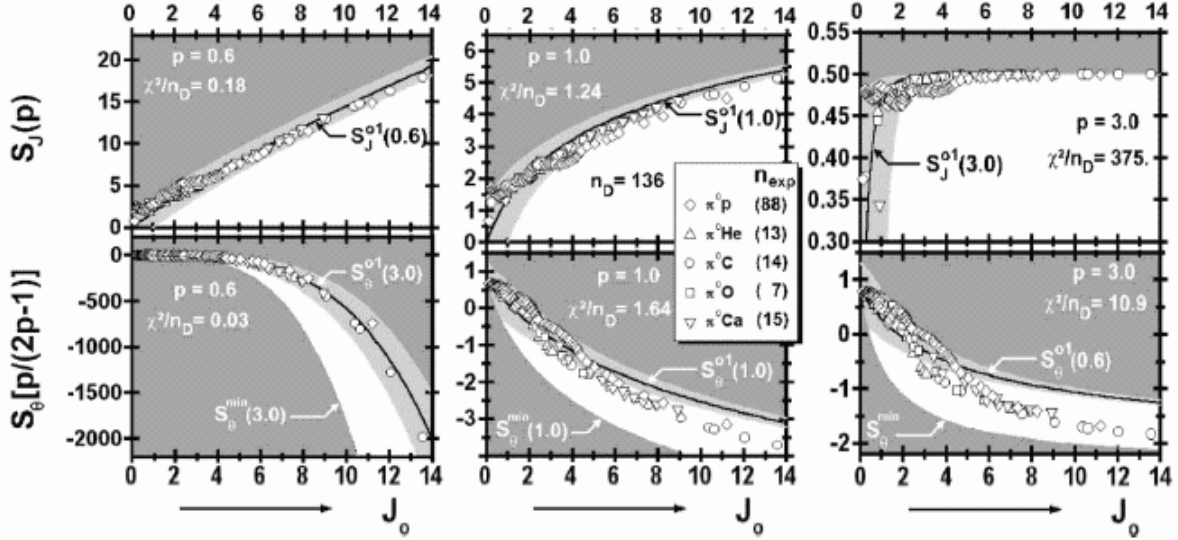

Fig 8. The experimental values of the scattering entropies $S_\theta(q)$ and $S_L(q)$, are compared with the theoretical optimal state predictions (full curve) for pion-nucleus (He, C, O, and Ca) scatterings (see the paper [13] for details)

An experimental verification of the correlation (49) is illustrated in Figs. 8-9. In Fig 9 the experimental values of the Tsallis-like entropies $S_\theta(q), S_J(p)$ and $\overline{S}_{J\theta}(p,q)$, calculated by using Eqs. (29), (31) and (35) p=0.6 and q=3 and experimental pion-nucleon [20] and kaon-nucleon [32] phase shifts, are compared with the theoretical optimal state predictions (full curves) for pion-nucleon and kaon-nucleon scatterings. The grey regions from Figs. 8-10 are obtained by assuming a minimum error of $\Delta J_o = \pm 1$ in the estimation of the optimal angular momentum from the experimental data.

## 8. Limited entropic uncertainty principle [15]

In the paper [14] by introduction of the nonextensivity conjugated entropy $\overline{S}_{J\theta}(p,q)$ we proved the following new *nonextensive conjugated entropic uncertainty relations* (NC-EUR) as well as *new nonextensive conjugated entropic uncertainty bands* (NC-EUB). If $\sigma_{el}$ and $\frac{d\sigma}{d\Omega}(1)$ are known from experiments then the nonextensivity conjugated entropy $\overline{S}_{J\theta}(p,q)$, defined by Eqs.(35)-(38), must obey the following entropic bands:

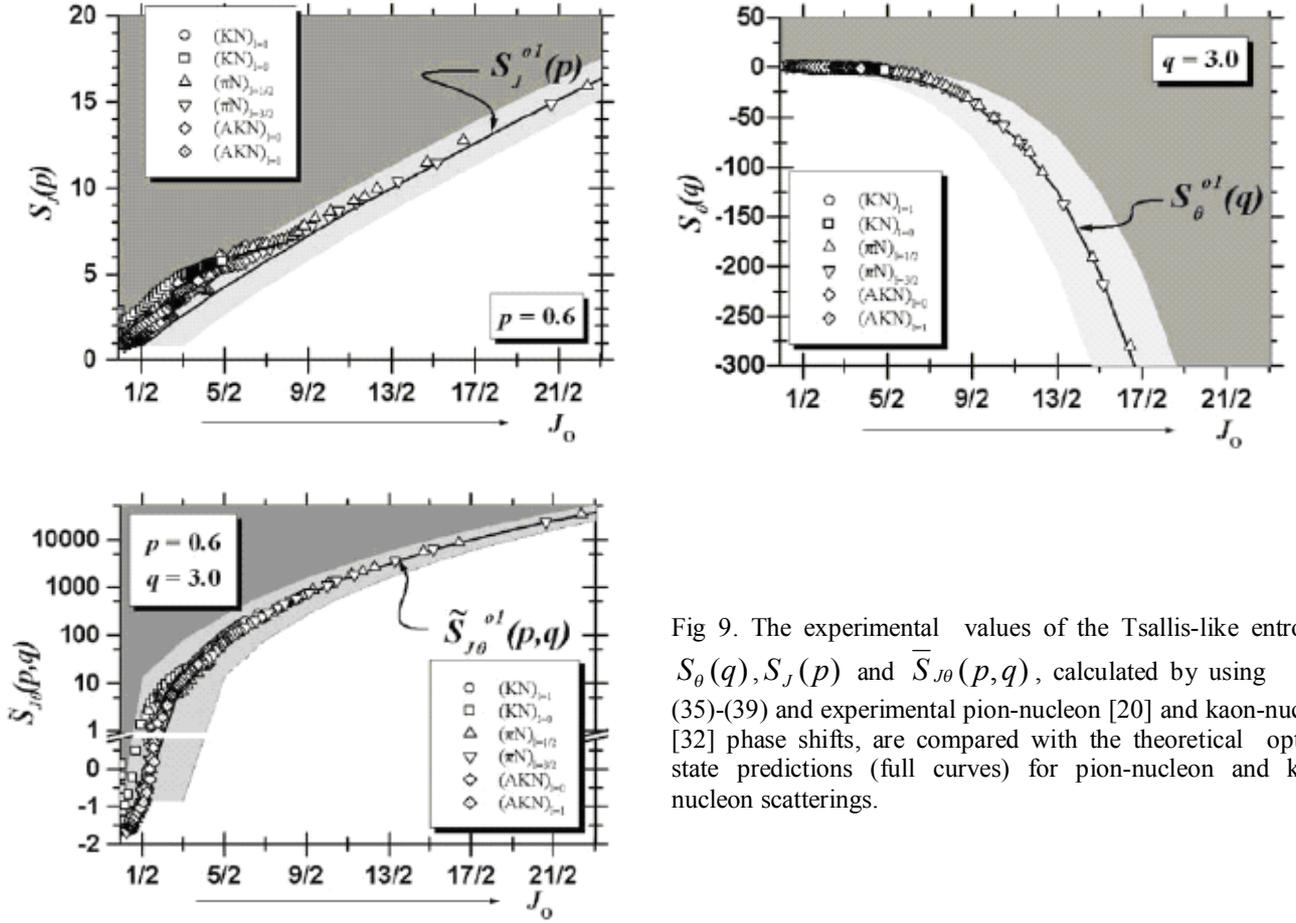

Fig 9. The experimental values of the Tsallis-like entropies $S_\theta(q)$, $S_J(p)$ and $\overline{S}_{J\theta}(p,q)$, calculated by using Eqs. (35)-(39) and experimental pion-nucleon [20] and kaon-nucleon [32] phase shifts, are compared with the theoretical optimal state predictions (full curves) for pion-nucleon and kaon-nucleon scatterings.

$$\overline{S}_{J\theta}^{\min}(p,q) \leq \overline{S}_{J\theta}(p,q) \leq \overline{S}_{J\theta}^{\max}(p,q) \qquad \text{q=p/(2p-1)} \quad \text{(NC-EUB)} \qquad (57)$$

$$\overline{S}_{J\theta}^{\min}(p,q) = \frac{1}{p-1}\left[1-[2]^{1-p}\int_{-1}^{1}dx\left(\frac{K_{\frac{1}{2}\frac{1}{2}}^{2}(x,1)}{K_{\frac{1}{2}\frac{1}{2}}(1,1)}\right)^{q}\right] \text{(NC-EUR)} \qquad (58)$$

$$\overline{S}_{J\theta}^{\max}(p,q) = \overline{S}_{J\theta}^{o1}(p,q) = \frac{1}{p-1}\left[1-[2K_{\frac{1}{2}\frac{1}{2}}(1,1)]^{1-p}\int_{-1}^{1}dx\left(\frac{K_{\frac{1}{2}\frac{1}{2}}^{2}(x,1)}{K_{\frac{1}{2}\frac{1}{2}}(1,1)}\right)^{q}\right] \text{PMD-SQS-prediction} \qquad (59)$$

for any $p \in (1/2,\infty)$ and q=p/(2p-1).

For numerical investigation of the optimal NC-EUB (57) we calculated nonextensive scattering entropies by using the available phase shifts [20] for all charge $\pi P \to \pi P$-channels.

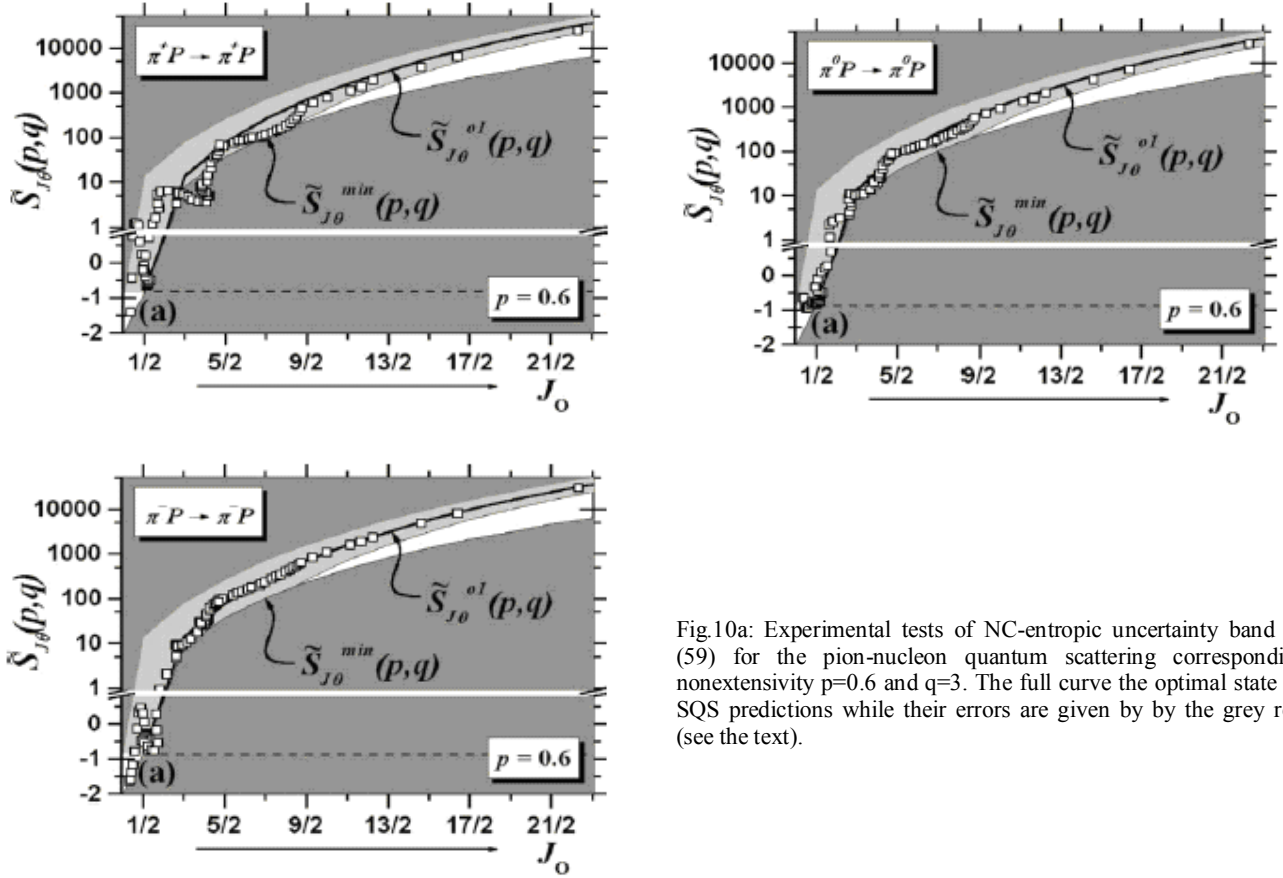

Fig.10a: Experimental tests of NC-entropic uncertainty band (57)-(59) for the pion-nucleon quantum scattering corresponding to nonextensivity p=0.6 and q=3. The full curve the optimal state PMD-SQS predictions while their errors are given by by the grey regions (see the text).

So, in Figs. 10a,b,c we presented the first experimental test of the uncertainty entropic inequalities (51)-(53) in pion-nucleon scattering for three kinds of nonextensivity conjugated statistics: (a) [p=2/3, q=2]; (b) [p=1,q=1] and (c) [p=2, q=2/3], So, the experimental results for $\overline{S}_{J\theta}[p, p/(2p-1)]$ are compared with the PMD=SQS predictions (full curve). Hence, by the numerical results displayed in Fig. 10a,b,c, we obtained not only consistent experimental tests of the optimal NC-EUR and NC-EUB, but also strong experimental evidences for the principle of minimum distance in space of quantum states (PMD-SQS). The grey regions from Figs. 10 are obtained by assuming a minimum error of $\Delta J_o = \pm 1$ in the estimation of the optimal angular momentum $J_o = half \, \mathrm{int} eger \left[ \dfrac{4\pi}{\sigma_{el}} \dfrac{d\sigma}{d\Omega}(1) + 1/4 \right]^{1/2} - 1$ from the experimental data [20].

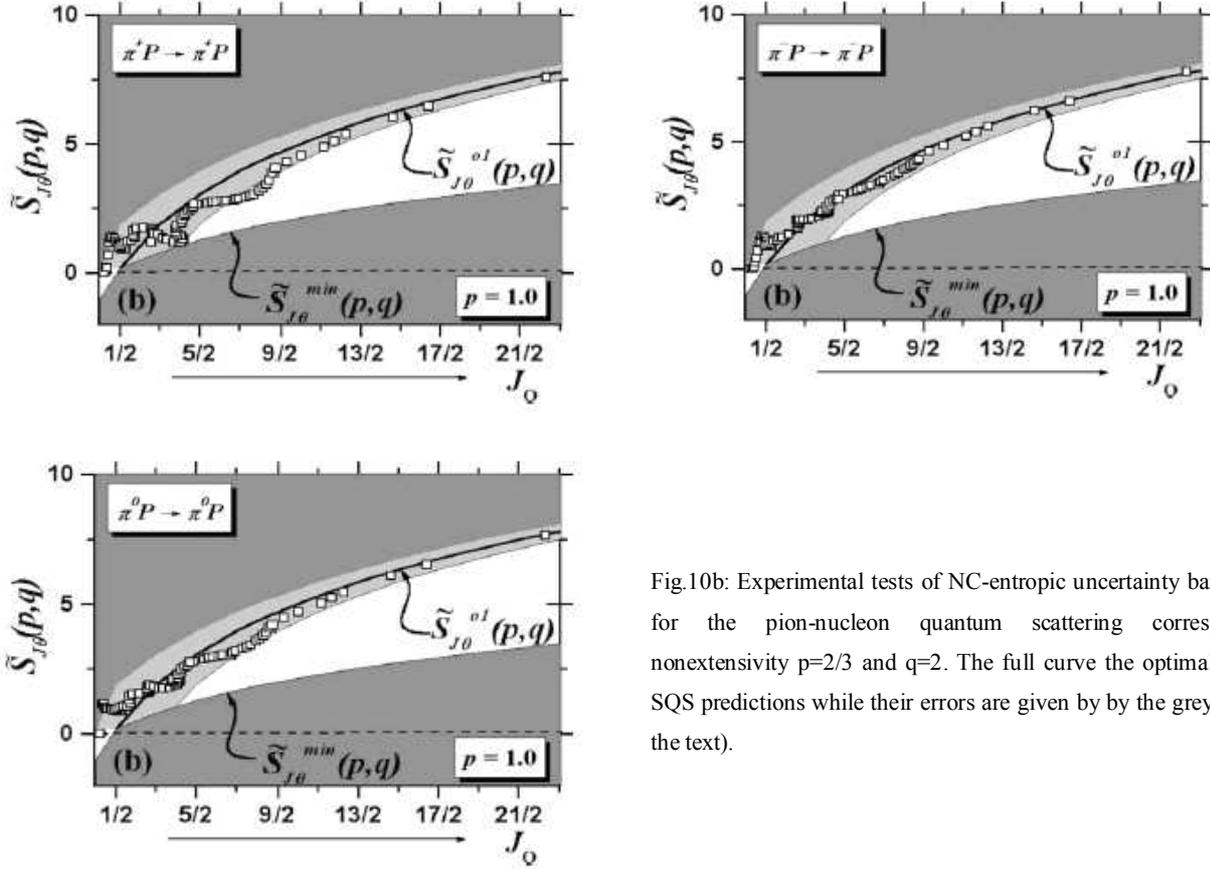

Fig.10b: Experimental tests of NC-entropic uncertainty band (57)-(59) for the pion-nucleon quantum scattering corresponding to nonextensivity p=2/3 and q=2. The full curve the optimal state PMD-SQS predictions while their errors are given by by the grey regions (see the text).

From Figs. 10a,b,c, we see that the experimental data on the nonextensivity conjugated entropies $\bar{S}_{J0}[p, p/(2p-1)]$ are in excellent agreement (CL>99%) with the PMD-SQS optimal predictions if the nonextensivities p and q of the scattering entropies $S_J(p)$ and $S(q)$ are correlated via Riesz-Thorin relation $\left[\dfrac{1}{2p} + \dfrac{1}{2q} = 1\right]$ [or q=p/(2p-1)]. So, the best fit is obtained for the conjugate pairs [p,q=p/(2p-1)] with the values of nonextensivity p in the range $1/2 \le p \le 2/3$. In fact a significant departure from PMD-SQS-optimality is observed only for conjugate pairs with $p \ge 2$ and $1/2 \le q \le 2/3$ where $\chi^2/n_D > 500$.

The results obtained in Fig. 10 for $\bar{S}_{J0}[p, p/(2p-1)]$ can be compared with those for the entropies $S_{J0}(p,q)$, for p=q presented in Fig.4 in Ref.[15]. Then, we see that the values of the usual $S_{J0}(p,q)$ are far from optimal values $S_{J0}^{o1}(p,q)$, p=q, and in some cases violate the entropic uncertainty band.

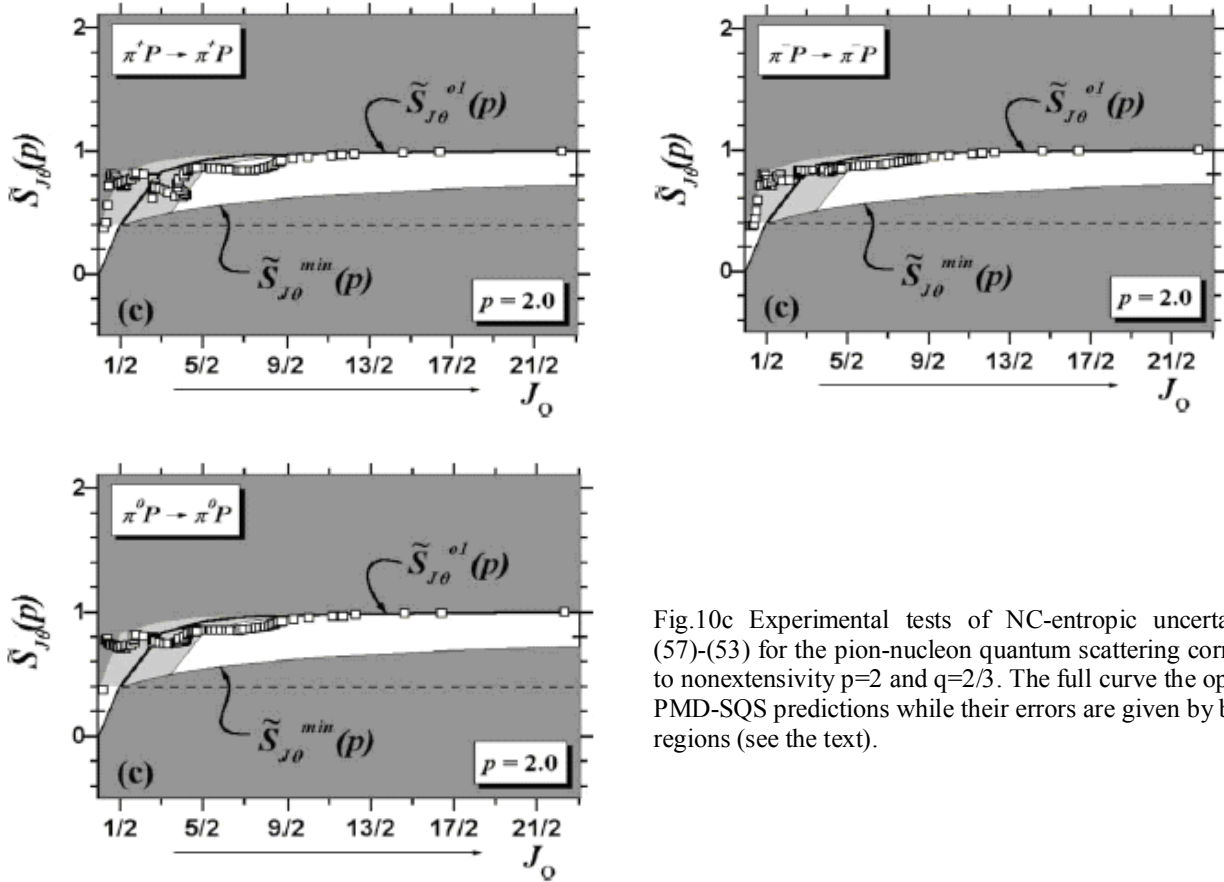

Fig.10c Experimental tests of NC-entropic uncertainty band (57)-(53) for the pion-nucleon quantum scattering corresponding to nonextensivity p=2 and q=2/3. The full curve the optimal state PMD-SQS predictions while their errors are given by by the grey regions (see the text).

These results alow us to conclude thar the new entropy $\overline{S}_{J\theta}[p, p/(2p-1)]$ is the best candidate for description of entropic uncertainty relations when the nonextensivity conjugation is taken into account.

## 9. Conclusions and Outlook

The significant results obtained in Ref. [1-16], can be summarized as follows:

[i] A description of quantum scattering via principle of minimum distance in space of states.[1]

[ii] Optimal state description of hadron-hadron via RKHS methods.

[iii] A proof of two important physical laws: Scaling and S-channel helicity conservation in hadron-hadron scattering.

[iv] A proof of Limited Entropic Uncertainty Principle (LEU-Principle) as new principle in quantum physics.

[v] Experimental evidences for optimal conjugated nonextensive statistics in hadronic scatterings.

[vi] Introduction of a new nonextensive quantum entropy and discovery of equilibrium of the quantum hadronic states.

[vii] The strong experimental evidence obtained here for the nonextensive statistical behavior} of the $(J, \theta)-$ quantum scatterings states in the pion-nucleon, kaon-nucleon and antikaon-nucleon scatterings can be interpreted as an indirect manifestation the presence of the quarks and gluons as fundamental constituents of the scattering system having the strong-coupling long-range regime required by the Quantum Chromodynamics.

The theory of diffraction obtained via PMD-SQS in our paper [1] is free of any paradox of the old quantum physics in the sense that diffraction and interference are universal phenomena independent of particle-wave duality interpretation, possessing not only universal character but also "scaling" properties. Such universal diffraction or universal interference phenomenon can occur for any massive particle. Therefore a continuation of the papers [1-17] is needed since the theory based on the PMD-SQS-optimality includes the heart of quantum phenomena.

**Universal PMD-SQS-optimal bound states**. Special investigations will be conducted for the development of a theory of the PMD-SQS-optimal bound states. We prove that the PMD-SQS-optimal "universal atomic levels" are all of form: $E_n = R/n^2$ , where $R = m_a + m_N - M_A$ is a Rydberg-like energy while optimal principal quantum number is $n = (L_o + 1)$. We will determine the analytic expression of the PMD-SQS-optimal "atomic" amplitudes in uni-directional, as well as, in bidirectional optimal models. For atomic transitions, we expect to prove that the optimal spectral series $\omega = R \left[ \dfrac{1}{n_o^2} - \dfrac{1}{m_o^2} \right]$ are direct consequences of the energy conservation in that transition. Such spectral series are just those observed from the atomic experimental data. The construction of a possible optimal Mendeleyev-like Table will also be investigated. We try to get an answer to the following important questions: Can atoms be completely described by the PMD-SQS-optimum principle without Coulomb potential and without using Schrödinger or Dirac equation? Also, similar problems of optimality can be formulated just in the same way for the nuclear structures, chemical structures or even for the gravitational bound states structures.

**Universal optimal resonances**: Special PMD-SQS-optimality investigations also will be dedicated to the development of a theory of the PMD-SQS-optimal resonances. Here, many important results are expected to be proved. (i)<u>The very high widths </u>for the optimal resonance, where: $\Gamma_n^o = \Gamma_1 n$, $n = L_o + 1$. (ii) A definite correlation between "positions" and "widths" of "elementary" resonances which are constituents of the optimal resonances, (iii) A "diffractive" character of the angular distributions, (iv) A total intensity proportional with $(L_o+1)^2$. Moreover, our investigation will be conducted to get from the available data the experimental evidences for the optimal resonances in hadron-hadron scattering as well as in hadron-nucleus scattering. We must underline that not only so called "diffractive (Morrison) resonances" ($A_1, A_2, etc.$) are candidates for to be optimal resonances, but also, the "dual diffractive resonances" discovered by us and published in the paper [17].

Finally, we remark that the physics based on the PMD-SQS minimum principle is a new branch of the quantum physics interconnecting, in a creative way, quantum physics, functional analysis, group theory, as well as quantum information theory. All these new ideas and theoretical developments will allow us to look at natural phenomena in a radically new and original way, eventually leading to unifying concepts independently of the detailed structures of systems not only in physics, but also in other sciences such as: Astrophysics, Biology, Psychology, Genetics, Economy, etc.